\documentclass[useAMS,usenatbib]{mn2e}
\usepackage{graphicx}
\usepackage{graphicx}
   \title{New white dwarf stars in the Sloan Digital Sky Survey Data Release 10}

   \author[Kepler et al.]{S. O. Kepler$^{1}$\thanks{kepler@if.ufrgs.br},
		I. Pelisoli$^{1}$,
		D. Koester$^{2}$,
		G. Ourique$^{1}$,
		S. J. Kleinman$^{3}$,
		A. D. Romero$^{1}$,
\newauthor
		A. Nitta$^{3}$,
		D. J. Eisenstein$^4$,
                J.E.S. Costa$^{1}$,
                B. K\"ulebi$^{5,6}$,
S. Jordan$^7$,
	P. Dufour$^8$,
\newauthor
Paolo Giommi$^9$,
                and
Alberto Rebassa-Mansergas$^{10}$\\
$^{1}$Instituto de F\'{\i}sica, Universidade Federal do Rio Grande do Sul,
              91501-900  Porto-Alegre, RS, Brazil\\
$^{2}$Institut f\"ur Theoretische Physik und Astrophysik, Universit\"at Kiel, 24098 Kiel, Germany\\
$^{3}$Gemini Observatory, Hilo, Hawaii, 96720, USA\\
$^4$Harvard Smithsonian Center for Astrophysics, 60 Garden St., MS \#20, Cambridge, MA 02138, USA\\
$^{5}$Institut de Ci\`encies de L'Espai, Universitat Aut\`onoma de Barcelona, 08193 Bellaterra, Spain\\
$^{6}$Institute for Space Studies of Catalonia, c/Gran Capit\`a 2-4, Edif. Nexus 104, 08034 Barcelona, Spain\\
$^8$Astronomisches Rechen-Institut, Zentrum f\"ur Astronomie der Universit\"at Heidelberg, M\"onchhofstr. 12-14, \\ D-69120 Heidelberg, Germany\\
$^8$D\'epartement de Physique, Universit\'e de Montr\'eal, C. P. 6128, Succ. Centre-Ville, Montr\'eal, Qu\'ebec H3C 3J7, Canada\\
$^9$ASDC Agenzia Spaziale Italiana, Via del Politecnico snc, 00133 Rome, Italy\\
$^{10}$Kavli Institute for Astronomy and Astrophysics, Peking University, Beijing 100871, People's Republic of China
}
\begin{document}
\date{Accepted 2014 November 8.  Received 2014 November 8; in original form 2014 July 28}

\pagerange{\pageref{firstpage}--\pageref{lastpage}} \pubyear{2014}
   \maketitle

\label{firstpage}
  \begin{abstract}

We report the discovery of $9\,089$ new spectroscopically 
confirmed white dwarfs and subdwarfs in the 
Sloan Digital Sky Survey Data Release 10. 
We obtain $T_\mathrm{eff}$, log g and mass for hydrogen atmosphere white dwarf stars 
(DAs) and helium
atmosphere white dwarf stars (DBs), and 
estimate the calcium/helium abundances for the white dwarf stars with metallic lines
(DZs) and carbon/helium for carbon dominated spectra DQs. 
We found 1 central star of a planetary nebula, 2 new oxygen spectra on helium atmosphere white dwarfs, 71 DQs, 42 hot DO/PG1159s, 
171 white dwarf+main sequence star binaries,
206 magnetic DAHs, 
327 continuum dominated DCs, 
397 metal polluted white dwarfs,
450 helium dominated white dwarfs, 
647 subdwarfs and 6888 new hydrogen dominated white dwarf stars.

\end{abstract}


\section{Introduction}
White dwarf stars are the end product of evolution of all stars with progenitor masses below
8--10.5~$M\odot$, depending on metallicity \citep{Doherty14}.
The first full white dwarf catalogue from Sloan Digital Sky Survey data
\citep{kle04} was based on SDSS Data Release 1 \citep[DR1,][]{dr1}.
Using data from the SDSS Data Release 4 \citep[DR4,][]{dr4}, \citet{eis06}
reported over 9\,000 spectroscopically-confirmed white dwarfs stars.
In the latest white dwarf catalogue based on the 
SDSS Data Release~7, \citet[DR7,][]{dr7} classified the spectra
of 19\,713 white dwarf stars, of which 12\,831 were
hydrogen atmosphere white dwarf stars (DAs) and 922 helium
atmosphere white dwarf stars (DBs), 
including the (re)analysis of stars from
previous releases.

The SDSS data massively increased the number of known white dwarfs, 
which had an enormous impact on the knowledge of these stars. 
However, target selection considerations of the original SDSS implied that white 
dwarf selection for spectroscopy was incomplete. 
White dwarf stars were targeted in SDSS-III's \citep{Eisenstein11}
Baryon Oscillation Spectroscopic Survey (BOSS)
by the ancillary project
White Dwarf and Hot Subdwarfs \citep{Dawson13}.
By DR10 the ancillary target programme obtained the spectra of $3\,104$ colour selected white dwarf candidates
that were missed by prior SDSS spectroscopic surveys.
Here, we report on our full search for new white dwarfs from the SDSS Data Release 10 \citep{dr10}.
Our catalogue does not include stars from the earlier
catalogues.

\section{Target selection}
SDSS multi-colour imaging separates hot white dwarf and subdwarf stars 
from the bulk of the stellar and quasar loci in colour-colour space 
\citep{Harris03}. Special target classes in SDSS produced the world's 
largest spectroscopic samples of white dwarfs.
However, much of SDSS white dwarf targeting required that the objects be 
unblended, which caused many brighter white dwarfs to be skipped 
(for a detailed discussion, see Section 5.6 of Eisenstein et al. 2006). 
The BOSS ancillary targeting programme \citep{Dawson13} relaxed this requirement and imposed colour
cuts to focus on warm and hot white dwarfs. 
Importantly, the BOSS spectral range extends further into the UV, 
allowing full coverage of the Balmer lines.

The targeted white dwarfs were required to be point sources with clean photometry, and to have
USNO-B Catalog counterparts \citep{Monet03}. They were also restricted to regions inside the DR7 imaging footprint 
and required to have colours within the ranges
$g < 19.2$,
$(u-r) < 0.4$,
$-1 < (u-g) < 0.3$,
$-1 < (g-r) < 0.5$, and to have low
Galactic extinction $A_r < 0.5$~mag.
Additionally, targets that did not have 
$(u-r) < -0.1$ and $(g-r) < -0.1$ were required to have USNO proper motions larger than 
2 arcsec per century.
Objects satisfying the selection criteria that had not been
observed previously by the SDSS were denoted by the WHITEDWARF\_NEW target flag, 
while those with prior SDSS spectra are assigned the WHITEDWARF\_SDSS flag. 
Some of the latter were re-observed with BOSS in order to obtain the extended wavelength coverage that the BOSS spectrograph offers.

The colour selection used includes DA stars with temperatures 
above $\sim 14\,000$~K, helium atmosphere white dwarfs above $\sim 8\,000$~K, 
as well as many rarer classes of white dwarfs. 
Hot subdwarfs (sdB and sdO) were targeted as well. 

SDSS-III BOSS \citep{Dawson13}
optical spectra extends from 3\,610~\AA\ to 10\,140~\AA, with spectral resolution 1560-2270 in the blue channel, and 1850-2650 in the red channel
\citep{Smee13}. The data
were reduced by the spectroscopic reduction pipeline of \citet{Bolton12}.

In addition to the 3104 targeted new white dwarf candidates, we selected the spectra of
any object classified by the {\small ELODIE} pipeline \citep{Bolton12} as a white dwarf,
which returned 27\,372 spectra.
Our general colour selection from \citet{dr7} returned
265\,633 spectra of which 42\,154 had already been examined in \citet{dr7}. From these,
we examined the spectra of additional
$12\,340$ with $g\leq 20$ and no Quasi Stellar Object (QSO) flag.
On top of those, we analyzed another 35\,000 spectra selected on the SDSS 3\,358\,200 optical spectra reported by DR10
with an automated search algorithm 
loosely based on
\citet{Si14}, estimating local averages at 75 pre-selected wavelengths sampling white dwarf strong lines,
and found another 1010 stars.
Of the $3\,104$ objects targeted specifically as new white dwarf spectra by BOSS as an ancillary programme, 574 were not identified as white dwarfs or subdwarfs by us.
From the total inspected by eye, around 37\% of the selected spectra are in fact confirmed white dwarfs,
of which
20\% were already known.
\citet{dr7} reported
47\% of their colour selected sample are white dwarf stars.
Of the Ancillary Program 43 of WHITE\-DWARF\_SDSS already observed,
157 in 2432 colour selected stars are in fact quasars.
These fractions must be considered when using samples of photometrically
selected white dwarfs, for example, to calculate luminosity functions.

We applied automated selection techniques supplemented by complete,
consistent human identifications of each candidate white dwarf spectrum.  

\section{Data Analysis}
After visual identification of the spectra as a probable white dwarf, 
we first fitted the
optical spectra to DA and DB local thermodynamic equilibrium (LTE) grids of
synthetic non-magnetic spectra derived from model atmospheres
\citep{Koester10}.  
The DA model grid uses the
ML2$/\alpha=0.6$ approximation,
and for the DBs we use the ML2/$\alpha=1.25$ approximation,
to be consistent with \citet{dr7}.
Our DA grid extends up to $T_\mathrm{eff}=100\,000$~K,
considering \citet{Napiwotzki97} concluded
pure hydrogen atmospheres of DA white dwarfs are well represented by LTE calculations for effective temperatures up to 80\,000~K. 
Only when
traces of helium are present, 
non-local thermodynamic equilibrium (NLTE)
effects on the Balmer lines occur, down to effective temperatures of 40\,000~K, recommending the neglect of traces of helium in the LTE models for the analysis of DA white dwarfs. 
We fitted all candidate white
dwarf spectra and colours with the {\small AUTOFIT}
code described in \citet{kle04}, \citet{eis06} and \citet{dr7}.
{\small AUTOFIT} fits only clean DA and DB models, so it does not
recognize other types of white dwarf stars.  In addition to the best
fitting model parameters, it also outputs a goodness of fit estimate and several
quality control checks and flags for other features noted in the spectrum
or fit.
The automatic fits include SDSS imaging photometry and allow
for refluxing of the models by a low-order polynomial to incorporate
effects of unknown reddening and spectrophotometric flux calibration errors.  
In addition, we have also fitted the spectral lines and photometry
separately \citep{Koester10}, selecting between the hot and cool solutions
using photometry as an indicator.
We make use of the latest SDSS reductions for photometry and
spectroscopy.
The white dwarf model atmospheres in our {\small AUTOFIT} spectral fits,
providing reliable $\log{g}$ and $T_\mathrm{eff}$ determinations for each identified clean DA
and DB are the same used in \citet{dr7}. We use the word {\it clean} to identify spectra that
show only features of non-magnetic, non-mixed, DA or DB (and DBA) stars.

We did not further restrict our sample by magnitude.
The SDSS
spectra we classified as white dwarfs or subdwarfs have a 
g-band signal--to--noise ratio $91 \geq S/N(g) \geq 1$, with an average of 15.
The lowest S/N in the g-band occurs for
stars cooler than $7\,000$~K, but they have significant S/N in the red
part of the spectrum.

\subsection{Spectral Classification}

Because we are interested in obtaining accurate mass distributions for
our DA and DB stars, we were conservative in labelling a spectrum as a clean
DA or DB,  adding additional subtypes
and uncertainty notations if we saw signs of other elements, companions, or
magnetic fields in the spectra.  While some of our mixed white dwarf
subtypes would probably be identified as clean DAs or DBs with better
signal-to-noise spectra,  few of our identified clean DAs or DBs would
likely be found to have additional spectral features within our detection
limit.

We looked for the following features to aid in the
classification for each specified white dwarf subtype:

\begin{itemize}
\item Balmer lines --- normally broad and with a Balmer decrement
[DA but also DAB, DBA, DZA, and subdwarfs]
\item HeI $4\,471$\AA\ [DB, subdwarfs]
\item HeII $4\,686$\AA\ [DO, PG1159, subdwarfs]
\item C2 Swan band or atomic CI lines [DQ]
\item CaII H \& K  [DZ, DAZ]
\item CII $4\,367$\AA\ [HotDQ]
\item Zeeman splitting [magnetic white dwarfs]
\item featureless spectrum with significant proper motion [DC]
\item flux increasing in the red [binary, most probably M companion]
\item OI $6\,158$\AA\ [Dox]
\end{itemize}

We also found a group of stars to have a very
steep Balmer decrement (i.e. only a broad H$\alpha$ and sometimes
H$\beta$ is observed while the other lines are absent) that could not
be fit with a pure hydrogen grid, or indicated extremely high gravities. 
We find that these objects are
best explained as helium-rich DAs, and denote them DA-He.

We finally note that the white dwarf colour space also contains
many hot subdwarfs. It is difficult, just by looking at a spectrum,
to tell a low mass white dwarf from a subdwarf,
as they are both dominated by hydrogen lines and the
small differences in surface gravity cannot be spotted by
visual inspection alone. We therefore extended the model grid
to $\log g=4.5$ to separate 
white dwarfs, hot subdwarfs and
main sequence stars
(see section \ref{section:masses} and \ref{section:sub}),
but the differences in the line widths for DAs cooler than $\simeq 8000$~K
and hotter than $\simeq 30\,000$~K are minor with changing gravity.

\subsection{Classification Results}
Table~\ref{tb:ids} lists the number of each type of white dwarf star we
identified.

\begin{table}
\begin{tabular}{rl}
{No. of Stars} &  {Type}\\
$6\,244$ &  DA$^a$\\
450 & DB$^b$\\
397 & DZ\\
327 & DC\\
234 & DAB\\
206 & DAH\\
247 & DAZ\\
171 & WD+MS$^c$\\
71 & DQ\\
62  & DAO\\
42 & DO/PG~1159\\
9 & DA-He\\
7 & DBZ\\
2 & Dox\\
1 & CSPN\\
636 & sd$^d$\\
\end{tabular}
\caption{\label{tb:ids}Numbers of newly identified white dwarf by type.}
$^a${Pure DAs.}
$^b${Include DBAs.}
$^{c}${These spectra show both a white dwarf star and a
companion, non-white dwarf spectrum, usually a main sequence M star.}
$^{d}${These are sdBs and sdOs subdwarf star spectra.}
\end{table}

The hot helium dominated spectra of PG~1159 and DO stars with S/N$\geq 10$ included in our Table~\ref{dados} were fitted with NLTE models and reported by
\citet{Werner14} and \citet{Reindl14}.

\section{Results}
\subsection{Oxygen Spectra}
\begin{figure}
   \centering
   \includegraphics[width=0.5\textwidth]{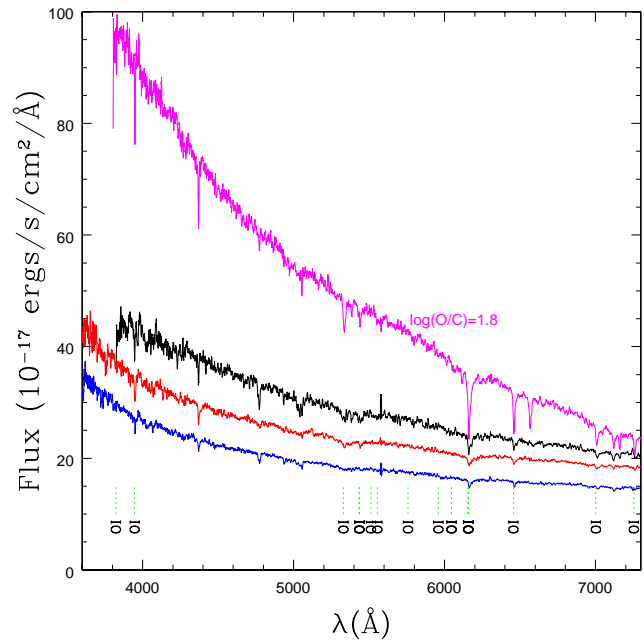}
      \caption{
Spectra of the prototype Oxygen dominated atmosphere white dwarf 
SDSS~J110239.69+205439.38=2488-54149-0167 at the top, 
SDSS~J142342.64+572949.3=2547-53917-0283, and the two new 
SDSS~J123807.42+374322.4=3969-55307-0772, and 
SDSS~J152309.05+015138.3=4011-55635-0072.
              }
         \label{dox}
   \end{figure}

Fig.~\ref{dox} shows the spectra of SDSS~J123807.42+374322.4, with $222 \pm 5$ mas/yr proper motion, $g=18.94\pm 0.01$,
and SDSS~J152309.05+015138.3, with $167 \pm 14$ mas/yr proper motion, $g=19.37\pm 0.01$,
have spectra dominated by OI lines,
that appear much stronger than the
C I lines, which can only be explained as being almost naked ONe cores, like
SDSS~J110239.69+205439.38, discovered by \citet{Boris10}.

Theoretically, stars with initial
masses $7 M_\odot \leq  M \leq  10 M_\odot$ 
will reach sufficiently high core temperatures to proceed to carbon burning,
and produce either oxygen-neon (ONe) core white dwarfs, or undergo a core-collapse supernova (SNII)
via electron capture on the products of carbon burning \citep{Nomoto84, Doherty14}. The exact
outcome of stellar evolution in this mass range depends critically on the detailed understanding of
the nuclear reaction rates involved, mass--loss and on the efficiency of convective mixing in the stellar
cores \citep[e.g.][]{Straniero03}. 
Observational constraints on stellar models 
come from analysis of SNII progenitors \citep{Valenti09,Smartt09a,Smartt09b},
suggest a lower limit on the progenitor masses of 
$8 \pm 1 M\odot$, but depends on progenitor metallicity.

\citet{Denissenkov13,Denissenkov14} propose the existence of hybrid C/O core white dwarfs with O-Ne envelopes
by stars where carbon is ignited off-centre but convective mixing prevents the carbon burning
to reach the centre.

\subsection{Magnetic Fields and Zeeman Splittings}
Similar to those reported for DR7 in \citet{dr7} and \citet{kepler13}, when examining each white dwarf candidate spectrum by eye,
we found 206 stars with Zeeman splittings
indicating magnetic fields above
2~MG --- the limit below which we cannot identify
since the line splitting is too small
for the SDSS spectral resolution. If
not identified as magnetic in origin, the spectra fittings of DA and DB models would have rendered
too high $\log g$ determinations due to magnetic broadening being misinterpreted as pressure broadening.
We also identified one DZH, in the same line as the DAZH identified by \citet{Kawka14}.
We estimated the mean fields following \citet{Kulebi}, from 2~MG to 310~MG.
We caution that stars with large fields are difficult to identify because any
field above around 30~MG, depending on effective temperature and
signal-to-noise, destroys the normal line sequences
and affect the colours significantly.
Additionally the fields above 100~MG represent the so-called
intermediate regime in which the magnetic white dwarf spectra have very few features,
save the stationary transitions which have similar wavelengths for a
distribution of magnetic fields.

\begin{figure}
   \centering
   \includegraphics[width=0.5\textwidth]{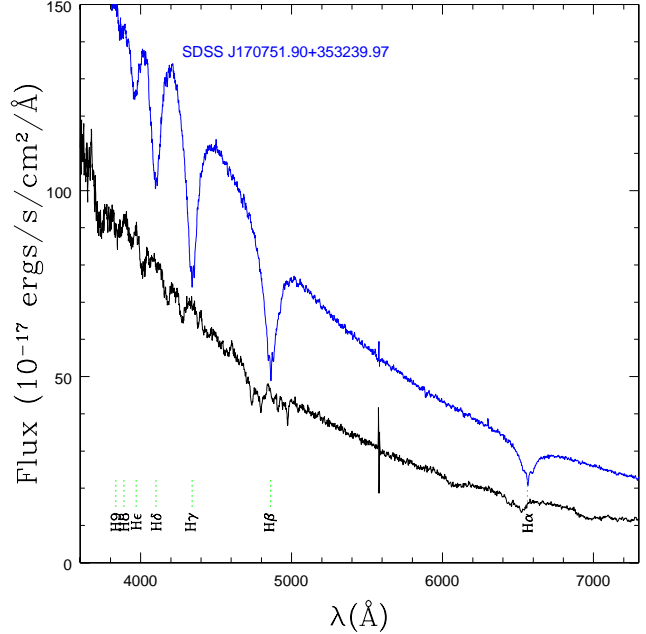}
      \caption{
Spectra of the DAH SDSS~J155708.04+041156.52, with
a mean magnetic field B=41~MG (bottom), 
and SDSS~J170751.90+353239.97, with B=2.7~MG (top).
              }
         \label{mag}
   \end{figure}

Fig.~\ref{mag} shows the S/N$_g$=42 and 58 spectra of the DAHs SDSS~J155708.04+041156.52, with
a mean magnetic field B=41~MG, and
SDSS~J170751.90+353239.97, also known as GD~359, with a mean field of 2.7~MG. 
For the magnetic fields 2--100~MG most relevant for magnetic white dwarfs, the quadratic
Zeeman effect is applicable. For this regime not only the spectral line
is divided into several components, the central component is also
displaced with respect to its non-magnetic value.

\subsection{Masses\label{section:masses}}

\citet{dr7} limited the white dwarf classification in the lower limit of surface gravity to $\log g=6.5$.
At the cool end of our sample, $\log g=6.5$ corresponds to a mass around $0.2~M\odot$, well below the single mass
evolution in the lifetime of the Universe, which corresponds to $\simeq 0.45~M\odot$,
depending on the progenitor metallicity.
All stars that form white dwarfs with masses below $\simeq 0.45~M\odot$ should be the byproduct of binary star evolution
involving interaction between the components, otherwise its lifetime on the main sequence would
be larger than the age of the Universe.

DA white dwarf stars with $\log g\leq  6.5$ and $T_\mathrm{eff} < 20\,000$~K are DA-ELM (Extreme Low Mass)
as found by
\citet{Brown10,Brown11,Brown12a,Brown12b,Brown13}.
\citet{Hermes12,Hermes13a,Hermes13b} found pulsations 
in five of these ELMs,
similar to
those of DAVs 
\citep{VanGrootel13}.

\begin{figure}
   \centering
   \includegraphics[width=0.5\textwidth]{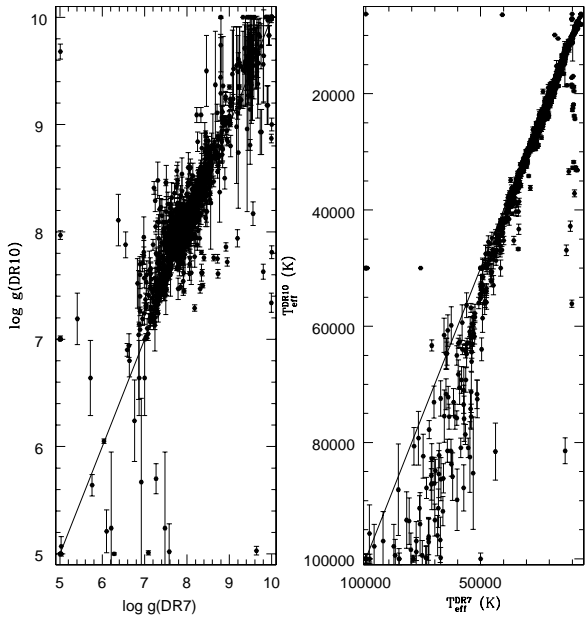}
      \caption{Surface gravities and effective temperature determinations for the same DA white dwarfs but
with spectra obtained with the SDSS spectrograph (DR7) and the BOSS spectrograph (DR10), with slightly larger
resolution and wavelength coverage.
         \label{diff}
              }
   \end{figure}

\begin{figure}
   \centering
   \includegraphics[width=0.5\textwidth]{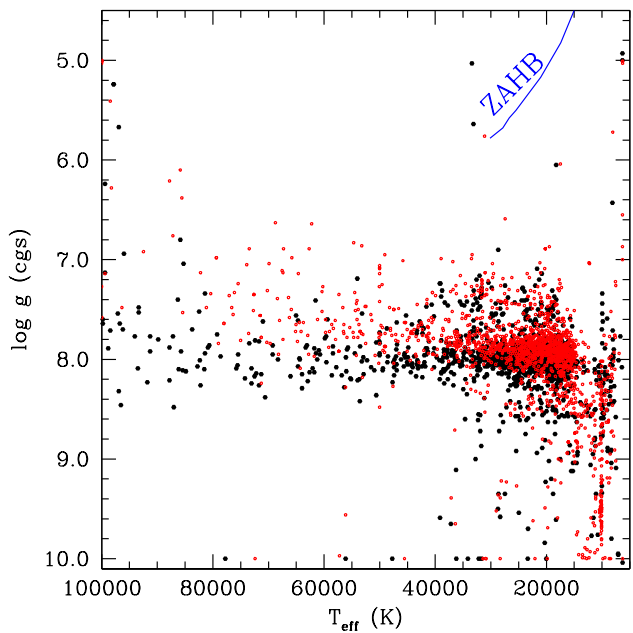}
      \caption{Surface gravities and effective temperature determinations for the same DA white dwarfs but
with spectra obtained with the SDSS spectrograph (open circles, DR7) and the BOSS spectrograph (filled circles, DR10), with slightly larger
resolution and wavelength coverage.
There is a systematic shift to higher gravities in the DR10 observations.
The line at the top right, marked ZAHB, shows the maximum $\log g$
a subdwarf can reach.
         \label{anc43}
              }
   \end{figure}

As a comparison of the DR7 and DR10 results, we plot in Figs.~\ref{diff} and \ref{anc43}
the $\log g$ and $T_\mathrm{eff}$ determinations of 1759 DAs observed
in Ancillary Program 43, i.e., for which new spectra with BOSS was
obtained of stars already observed with the SDSS spectrograph. 
Only 88 spectra have differences larger than 20\%,
mainly
at the high temperature end and around $17\,000~\mathrm{K} \geq T_\mathrm{eff} \geq 10\,000~\mathrm{K}$,
where there are two solutions in the spectral fittings,
but there is a systematic shift of the DR10 observations to higher surface gravity,
possibly caused by systematics in the flux calibrations \citep{Genest14}.

\begin{figure}
   \centering
   \includegraphics[width=0.45\textwidth]{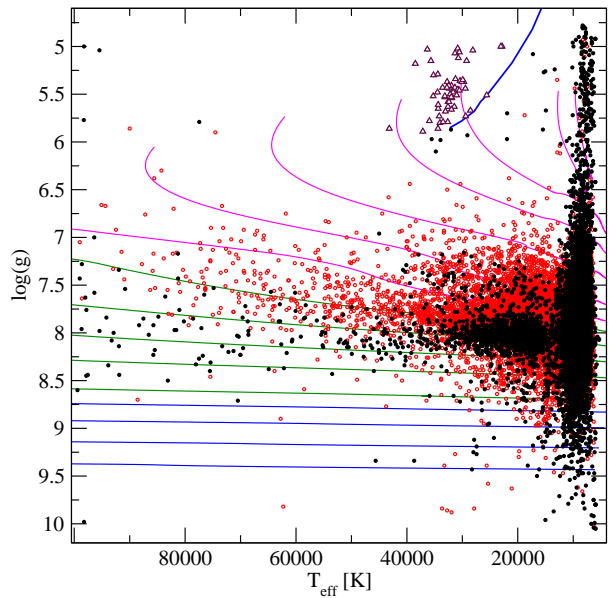}
      \caption{
Surface gravity ($\log g$) and effective temperature ($T_{\rm eff}$) estimated for the new DA white dwarf stars with spectra with S/N$_g\geq 10$, with atmospheric models corrected to three-dimensional convection
using the corrections reported in Tremblay et al. (2013) (black dots). 
The decrease in the number of DAs between $15\,000~\mathrm{K} \geq T_\mathrm{eff} \geq 11\,000$~K
is caused by selection effects -- the targeting for new white dwarfs selected only
DAs hotter than 14\,000˜K.
In open circles (red) we show the values for the DR7 DAs, but also corrected to three-dimensional convection
using the corrections reported in Tremblay et al. (2013).
It is evident that the surface gravities determined for the stars observed with the BOSS spectrograph, covering a larger wavelength range,
resulted in slightly higher surface gravities.
The zero-age horizontal branch (ZAHB) plotted was calculated specifically for this paper, with solar composition models.
It indicates the highest possible surface gravity for a subdwarf. Stars with 
$T_\mathrm{eff}\leq 45\,000$~K and
lower surface gravity 
than the ZAHB are sdBs,
indicated with triangles in the figure.
The low masses at the cool end are ELM white dwarfs, byproduct of binary evolution. The model lines show the ELM
are concentrated at low effective temperatures. They are present even when we restrict our spectra to the 852 DAs with S/N$\geq$ 25,
or even to the 125 with S/N$\geq$ 50.}
         \label{loggDA}
   \end{figure}

\begin{figure}
   \centering
   \includegraphics[width=0.5\textwidth]{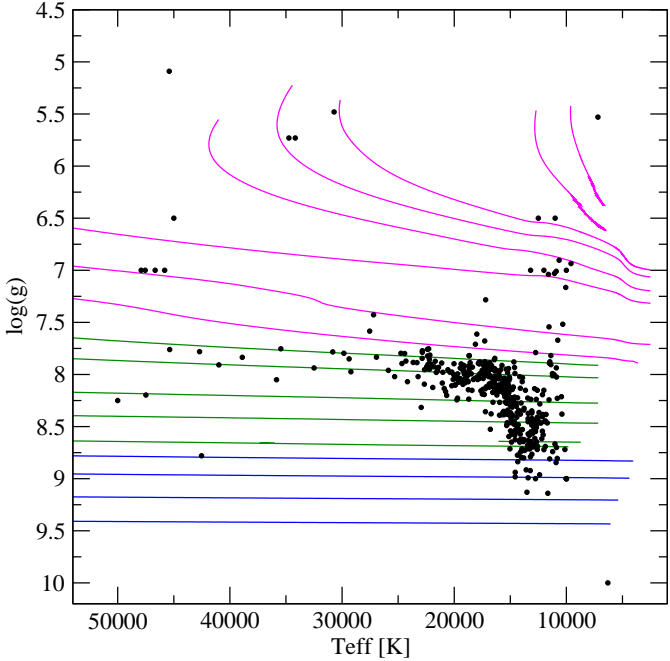}
      \caption{
Surface gravity ($\log g$) and effective temperature ($T_{\rm eff}$) estimated for the new DB white dwarf stars from models with convection described by ML2/$\alpha$=1.25 (black dots). 
Also plotted are full white dwarf evolutionary tracks. Carbon/oxygen sequences (green lines) correspond to He-rich atmosphere models with stellar mass ranging
from 0.515 (top) to 0.87$M{\odot}$, while helium- (magenta) and oxygen/neon-core (blue, bottom) sequences correspond to DA white dwarf models (see text for details).
We see a rise in $\log g$ for DBs cooler than $T_\mathrm{eff}=16\,000$~K,
similar to that reported by \citet{Kepler07} and also seen in
Fig.~21 of \citet{Bergeron11}, probably caused by incorrect neutral
broadening modelling.
              }
         \label{loggDB}
   \end{figure}

Fig.~\ref{loggDA} 
shows the surface gravity ($\log g$, in cgs units)
as a function
of the effective temperature ($T_\mathrm{eff}$, in K), estimated for all
DAs  
corrected to  three--dimensional convection  models using the
corrections presented in
\citet{Tremblay13}, and 
Fig.~\ref{loggDB} for
DBs with ML2/$\alpha=1.25$.

We use the
mass--radius relations of \citet{Renedo10} and \citet{Romero14}
carbon--oxygen  DA  white dwarfs,  for  solar metallicities,   
to calculate the mass of our identified clean DA stars from the
$T_\mathrm{eff}$ and  $\log g$ values obtained from our fits,
corrected to 3D convection.
These
relations are based on full evolutionary calculations appropriate for
the study of hydrogen-rich DA  white dwarfs that take into account the
full evolution  of progenitor stars, from the  zero--age main sequence,
through  the hydrogen and helium central burning stages, thermally
pulsating and mass--loss in the asymptotic giant branch phase and finally the planetary
nebula domain. The stellar  mass values  for the  resulting sequences
range from 0.525 to 1.024 $M{\odot}$, covering  the stellar
mass range for C--O core DAs. For high--gravity white dwarf
stars, we employed the mass--radius relations for O--Ne core
white dwarfs given in \citet{Althaus05}
in the mass range
from 1.06 to 1.30 $M{\odot}$ with a step of 0.02 $M{\odot}$. For
the low--gravity white dwarf stars, we used the evolutionary
calculations of \citet{Althaus13} for helium--core white dwarfs with
stellar mass between 0.155 to 0.435 $M{\odot}$, supplemented by
sequences of 0.452 and 0.521 $M{\odot}$ calculated in \citet{Althaus09a}.

For DB white dwarfs stars, we relied on the evolutionary calculations
of hydrogen--deficient white dwarf stars with stellar mass between
0.515 and 0.870 $M{\odot}$ computed by \citet{Althaus09b}. These
sequences have  been derived from the born--again episode responsible
for the  hydrogen deficient white dwarfs. For  high-- and low--gravity DBs, we used
the O-Ne and helium evolutionary sequences,
described before.

To calculate reliable mass distributions, we selected only the best S/N
spectra with temperatures well fit by our models.  We find that reliable
classifications can be had from spectra with S/N $\geq$ 10.
We classified $6\,243$
spectra as clean DAs.
Of these DAs, $2\,074$ have a spectrum with S/N $\geq 10$, with a mean S/N=$25\pm 13$, and
$\langle \log g_{\mathrm{DA}} \rangle=7.937 \pm 0.012$.
Table~\ref{mass} present the mean masses for different signat-to-noise limits,
after correcting to 3D convection models.
\begin{table}
\begin{center}
\begin{tabular}{crcrc}
S/N$_g$&N&$\langle M_{\mathrm{DA}} \rangle $&N&$\langle M_{\mathrm{DA}} \rangle$\cr
&&&&$T_\mathrm{eff} \geq 10\,000$ K\cr
&&$(M_\odot)$&&$(M_\odot)$\cr
10&2074&$0.626 \pm 0.004$&1573&$0.659 \pm 0.003$\cr  
15&1659&$0.656 \pm 0.004$&1484&$0.662 \pm 0.003$\cr
25& 852&$0.669 \pm 0.005$& 821&$0.677 \pm 0.005$\cr
50& 125&$0.663 \pm 0.014$& 120&$0.679 \pm 0.011$\cr
\end{tabular}
\caption{Mean masses for DAs, corrected to 3D convection\label{mass}}
\end{center}
\end{table}
The mean masses estimated in this sample are larger than those obtained by
\citet{dr7},
which did not use the 3D correction, but now closer to the results of other determinations methods, as gravitational redshift 
\citep{Falcon10} and seismology \citep{Romero12},
and especially with the \citet{Gianninas11} sample of bright DAs
corrected  to 3D by
\citet{Tremblay13} which is centred at
$0.637~M\odot$.

\begin{figure}
   \centering
   \includegraphics[width=0.5\textwidth]{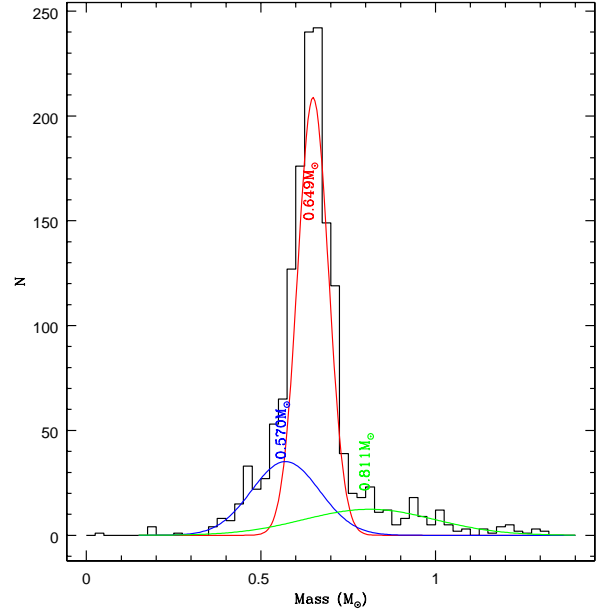}
      \caption{
Histogram for the mass distribution of S/N$\geq$10 and $T_\mathrm{eff}\geq 12\,000$~K DAs versus mass, for $\log g$ corrected to three-dimensional convection models using
the corrections reported in \citet{Tremblay13}.
Even though the mass histogram is not expected to have a Gaussian shape, we have decomposed it in three Gaussians, just as a guide:
$N=208.9*exp[-(M-0.649)^2/(2*0.044^2)]+
35.2*exp[-(M-0.570)^2/(2*0.097^2)]+
12.4*exp[-(M-0.811)^2/(2*0.187^2)]$.
As in \citet{Kepler07} and \citet{dr7}, we find a significant number of stars with masses below 0.45~$M\odot$,
product of interacting binary evolution.
              }
         \label{masshist}
   \end{figure}

Fig.~\ref{masshist} shows the mass histogram for the 1\,504 DAs with S/N $\geq$ 10 and $T_\mathrm{eff}\geq 12\,000$~K.

The spectra we classified as clean DBs belong to 381 stars.
373 of these have a spectral S/N $\geq 10$, with a mean S/N=$25\pm 11$.
Using this high S/N sample, we obtain
 $\langle \log g_\mathrm{DB} \rangle = 8.122 \pm 0.017$.
and a corresponding mean mass of
$\langle M_\mathrm{DB} \rangle = 0.696 \pm 0.010 M_\odot$,
similar to the results from gravitational redshift by
\citet{Falcon12} and \citet{Bergeron11},
and to
$\langle M_\mathrm{DB} \rangle = 0.685 \pm 0.013 M_\odot$
for the 191 DBs with S/N$\geq 15$ of \citet{dr7}.

\subsection{DZs}
9\% 
of white dwarfs cooler than $T_\mathrm{eff}=12\,000$~K in our sample show spectra contaminated by metals, probably
due to accretion of rocky material around the stars \citep[e.g.][]{Graham90, Jura03, Koester14}.
Calcium and Magnesium in general have the strongest lines for white dwarfs at these temperatures.

\begin{figure}
   \centering
   \includegraphics[width=0.5\textwidth]{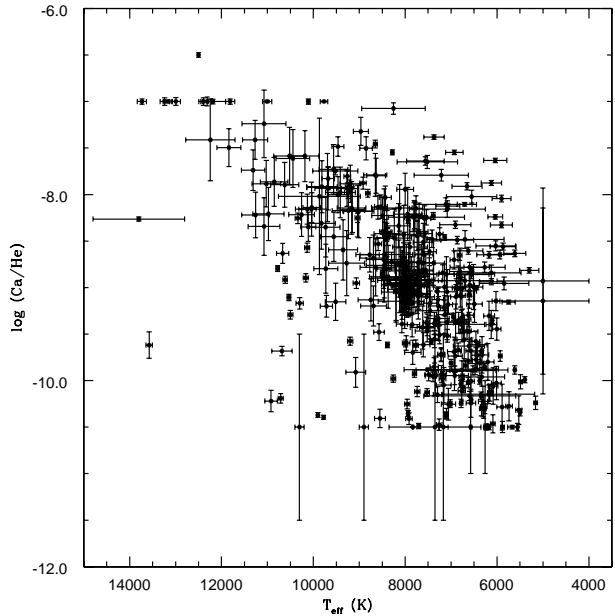}
      \caption{Calcium/Helium abundances estimated for DZs, assuming a bulk Earth composition for the accreted material.
              }
         \label{dzs}
   \end{figure}

We fitted the spectra of each of the 397 stars classified as DZs to
a grid of models with 
Mg, Ca and Fe ratios equal to the averages from the cool DZ in 
\citet{Koester11},
and Si added with the same abundance as Mg. This is fairly close to bulk Earth, except for the missing oxygen
\citep{Koester14}. The models have fixed
surface gravity at $\log g=8.0$ as it is not possible to estimate it
from the spectra. 
The absolute values for log Ca/He range from -7.25 to -10.50.
Fig.~\ref{dzs} shows the Calcium/Helium abundance for the 397 DZs identified.
There seems to be a decrease of Ca/He abundances at lower temperatures,
that could be explained by approximately the same accretion rate diluted by
increasing convection layer.

\subsection{DQs}
1\% 
of white dwarfs cooler than $T_\mathrm{eff}=12\,000$~K in our sample show spectra dominated by carbon lines, probably
due to dredge-up of carbon from the underlying carbon-oxygen core through the expanding He convection zone
\citep[e.g.][]{Koester82,Pelletier86,Koester06,Dufour07}.

\begin{figure}
   \centering
   \includegraphics[width=0.5\textwidth]{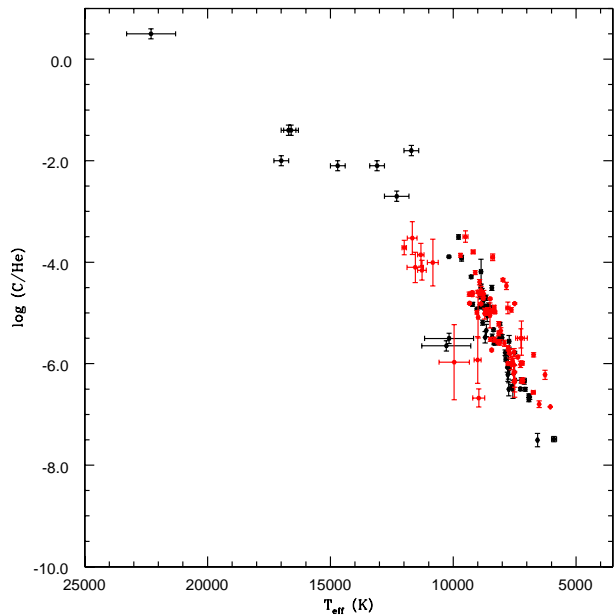}
      \caption{Carbon/Helium abundances estimated for DQs.
The increase with decreasing temperature comes from the increase in transparency and deepening convection zone.
The lighter (red) points are the results of our fits with the same models for the cool DQs in \citet{dr7}.
              }
         \label{dqs}
   \end{figure}

We fitted the spectra of the stars classified as cool DQs to
a grid of models reported by 
\citet{Koester06}.
The models have fixed
surface gravity at $\log g=8.0$ as it is not possible to estimate it
from the spectra. 
The absolute values for log C/He range from -8 to -4, and effective temperatures
from 13\,000~K to 4400~K.
The hotter DQs were fitted with the models
of \citet{Dufour11} and \citet{Dufour13}.
Fig.~\ref{dqs} shows the Carbon/Helium abundance for the 71 DQs identified.

\subsection{White dwarf-main sequence binaries}

We have identified 180 white dwarfs that are part of binary systems
containing main--sequence companions (WDMS binaries), of which 171 are
not included in the latest version of the SDSS WDMS binary catalogue
\citep{Rebassa13}.  
\citet{Li14} reports 227 DR9 DA+M pairs, 111 of which also found
by us independently. The
majority of our new systems contain a DA white dwarf and a M dwarf
secondary star (DA+M, see Table\,\ref{t-wdms}), and we measure their
stellar parameters following the decomposition/fitting routine
described in \citet{Rebassa07}. In a first step, a given
WDMS binary spectrum is fitted with a combination of M dwarf plus
white dwarf templates and the spectral type of the M dwarf is
determined. The best-fitting M dwarf template is then subtracted and the
residual white dwarf spectrum is fitted with the model grid of DA
white dwarfs of \citet{Koester10}. The fits to the normalized Balmer
lines are used to derive the white dwarf effective temperature and
surface gravity, and the fit to the entire spectrum (continuum plus
lines) is used to break the degeneracy between the hot and cold
solutions. The 3D corrections of \citet{Tremblay13} are applied to our
white dwarf parameter determinations.  
The stellar parameters are provided in Table~\ref{dados}.

\begin{table}
\centering
\caption{\label{t-wdms} Number of newly identified WDMS binary by types.}
\setlength{\tabcolsep}{0.8ex}
\begin{small}
\begin{tabular}{cccc}
\hline
\hline
   Classification & Number & Classification & Number\\
\hline
        DA+M  & 127  &       DB+M? &  1 \\
       DA+M:  & 12   &       DB:/M &  2 \\
        DA/K  & 2    &        DB:? &  1 \\
       DA:+M  & 2    &       DBA+M &  1 \\
        DB+?  & 1    &        DC+M &  2 \\
        DB+K  & 1    &       DC+M: &  2 \\
        DB+M  & 12   &        DA/F &  1 \\
\hline
\end{tabular}
\end{small}
\end{table}

\begin{figure}
\centering
\includegraphics[width=\columnwidth]{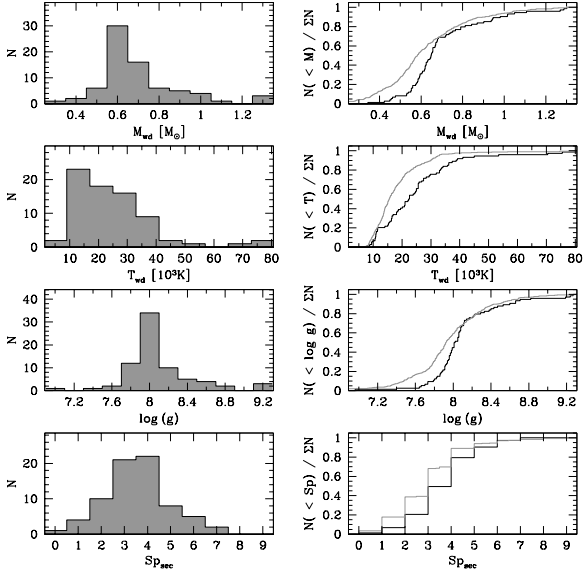}
\caption{\label{f-wdmshisto} Left: distribution of white dwarf mass
  (top), effective temperature (middle-top) and surface gravity
  (middle--bottom), and M dwarf spectral type (bottom) of new WDMS binaries
  identified in this work with spectra of SN$\geq10$. Right--hand panels:
  the cumulative parameter distributions, in the same order as in the
  left--hand columns (black lines). The cumulative parameter distributions
  of the DR8 WDMS binary catalogue are shown in gray (only parameters
  obtained from spectra of SN$\geq 10$ are considered).}
\end{figure}

76 of the new DA+M binaries in our sample have spectra with SN$\geq 10$, and for 74
(73) of them we derive reliable white dwarf parameters (M dwarf
spectral types) fitting their SDSS spectra (Table~\ref{dados}). The parameter
distributions are shown in Figure\,\ref{f-wdmshisto}, left--hand panels.  On
the right--hand panels of the same figure we show the cumulative parameter
distributions, which are compared to those obtained from the latest
version of the SDSS (DR8) WDMS binary catalogue
\citep{Rebassa13}. For a meaningful comparison, we apply
the 3D corrections of \citet{Tremblay13} to the DR8 data as well as
consider only systems with spectra of SN$\geq10$. Inspection of the
cumulative distributions of both studies reveals that white dwarfs in
the here presented WDMS binary sample are systematically hotter
(middle-top right--hand panel of Figure\,\ref{f-wdmshisto}), and that there
is a clear lack of low-mass white dwarfs in the DR10 sample as
compared to the DR8 catalogue (top and middle-bottom right--hand panels of
Figure\,\ref{f-wdmshisto}).  The fraction of early-type M dwarfs is
lower in the DR10 WDMS binary sample (Figure\,\ref{f-wdmshisto},
bottom panel).

\subsection{Subdwarfs\label{section:sub}}

Hot subdwarfs are core He burning stars.
Following \citet{Nemeth12,Drilling13,Nemeth14a,Nemeth14b},
we have classified stars with $\log g< 6.5$ as hot subdwarfs;
sdOs if He~II present and sdBs if not, between $45\,000~\mathrm{K} > T_\mathrm{eff} > 20\,000$~K.
\citet{Nemeth14a} and \citet{Rauch14} discuss how the He abundances
typical for sdB stars affect the NLTE atmosphere structure, 
and to a lower extent CNO and Fe abundances are
also important in deriving accurate temperatures and gravities. 
Our determinations of $T_\mathrm{eff}$ and $\log g$ do not include
NLTE effects or mixed compositions, so they serve only as a rough indicator.
Stars with $6.5 > \log g > 5.0$ and $T_\mathrm{eff} < 20\,000$~K were labelled ELM-DAs,
as they lay below the ZAHB
\citep{Dreizler02,Althaus13,Gianninas14}.
Of the 647 sds, we classified 262 sdOs and 375 sdBs.
We also identified $\simeq 200$ stars with $\log g < 5.2$ as either main sequence stars \citep{Arnold92},
even though their implied distance moduli would be larger than $m-M=16.5$
and most of them have galactic latitude $b>10\deg$,
or more likely as Extreme Low Mass white dwarfs \citet{Brown10,Brown11,Brown12a,Brown12b,Brown13}.

\subsection{DCs and BL~Lac}
\noindent
\begin{table*}
\begin{minipage}{\linewidth}
{\tiny
\begin{tabular}{lccccccccccccccc}
P-M-F          &SDSS J             &S/N&u              &g              &r              &i              &z              &ppm       &aro  &Radio\\
               &                   &   &               &               &               &               &               &(mas      &     &flux\\
               &                   &   &               &               &               &               &               &/yr)      &     &(mJy,\\
               &                   &   &               &               &               &               &               &          &     &1.4 GHz)\\
4394-55924-0742&021311.76-050102.56&21&20.42$\pm$00.06&19.95$\pm$00.02&19.54$\pm$00.02&19.20$\pm$00.02&18.87$\pm$00.04&    0     &0.500&32\\
4342-55531-0304&024519.63-025628.12&29&19.48$\pm$00.03&18.96$\pm$00.01&18.55$\pm$00.01&18.24$\pm$00.01&17.95$\pm$00.02&    0     &0.491&74\\
3802-55528-0818&080551.76+383538.04&17&21.64$\pm$00.11&20.99$\pm$00.04&20.38$\pm$00.03&19.94$\pm$00.03&19.59$\pm$00.07&    0     &0.488&13\\
4603-55999-0880&091141.72+424844.17&16&20.60$\pm$00.05&20.11$\pm$00.02&19.72$\pm$00.02&19.37$\pm$00.02&19.10$\pm$00.04& 7$\pm  3$&0.516&34\\
5399-55956-0224&122307.25+110038.28&23&19.89$\pm$00.04&19.61$\pm$00.01&19.39$\pm$00.01&19.13$\pm$00.02&19.00$\pm$00.04&    0     &0.250& 2\\
5484-56039-0716&150716.42+172102.89&15&19.74$\pm$00.03&19.40$\pm$00.01&19.03$\pm$00.01&18.77$\pm$00.01&18.64$\pm$00.03&$12\pm  6$&0.439&23\\
5188-55803-0712&163506.77+345852.24&14&20.76$\pm$00.06&20.29$\pm$00.02&19.83$\pm$00.02&19.49$\pm$00.02&19.15$\pm$00.05&$11\pm  4$&0.578&66\\
4318-55508-0464&220812.70+035304.61&17&19.85$\pm$00.04&19.40$\pm$00.01&18.92$\pm$00.01&18.59$\pm$00.01&18.29$\pm$00.03&$20\pm 17$&0.465&40\\
\end{tabular}
}
\caption{Featureless spectra objects that have radio emission consistent with BL~Lac classification. 
Four objects associated to some amount of proper motion,
SDSS~J091141.72+424844.17, SDSS˜J150716.42+172102.89, SDSS~J163506.77+345852.24 and SDSS~J220812.70+035304.61, 
are most likely BL~Lacs given their clearly non-thermal broad--band spectral energy distribution. In all these cases the measured proper motion is
not highly statistically significant and could have been affected by strong variability that is typical of BL~Lacs.
All these eight objects have H(21~cm) column densities larger than $10^{20}~\mathrm{cm}^{-2}$ and most
show variability in the Catalina Sky Survey \citep{Drake12}.  \label{BL}}
\end{minipage}
\end{table*}

Featureless optical spectra are typical of DC white dwarfs, but also from extragalactic BL~Lac objects. BL~Lac objects
are strong sources of radio, while non-interacting DCs are not. DCs, if bright enough to be detected in all images,
should have measurable proper motions, as they are cool faint objects.
To separate these objects, we searched for 1.4~GHz radio emission in the literature. BL~Lac objects have in general
a power-law spectra from optical to radio, with a radio to optical exponent larger than aro=0.250. We found 67/260 objects
we had classified by eye the spectrum as DCs are in fact known BL Lac objects. We also found 8 objects, listed in Table~\ref{BL},
with measured radio flux compatible with them being BL~Lac objects.

Table~\ref{tb:columns} lists the columns of data provided
in our electronic catalogue file, Table~\ref{dados}.
\begin{table*}
\caption{\label{tb:columns}Columns provided in data table, Table~\ref{dados}.}
\begin{tabular}{cll}
{Column No.} & {Heading} & {Description}
\\
1 & Name & SDSS object name (SDSS 2000J+) \\
2 & P-M-F & SDSS Plate number-Modified Julian date-Fiber\\
3 & SN\_g & SDSS g band signal to noise ratio \\
4 & u\_psf & SDSS u band PSF magnitude \\
5 & u\_err & SDSS u band uncertainty \\
6 & g\_psf & SDSS g band PSF magnitude \\
7 & g\_err & SDSS g band uncertainty \\
8 & r\_psf & SDSS r band PSF magnitude \\
9 & r\_err & SDSS r band uncertainty \\
10 & i\_psf & SDSS i band PSF magnitude \\
11 & i\_err & SDSS i band uncertainty \\
12 & z\_psf & SDSS z band PSF magnitude \\
13 & z\_err & SDSS z band uncertainty \\
14 & PM & SDSS proper motion (0.01\" yr$^{-1}$) \\
15 & T\_eff & $T_\mathrm{eff}$ \\
16 & T\_err & $T_\mathrm{eff}$ uncertainty \\
17 & log\_g & $\log{g}$ \\
18 & log\_gerr & $\log{g}$ uncertainty\\
19 & humanID & human classification\\
20 & T\_eff (3D) & $T_\mathrm{eff}$ for pure DAs and DBs or -log(Ca/He) for DZs or -log(C/He) for DQs\footnote{The temperatures and surface gravities are corrected to the three--dimensional
convection models of \citet{Tremblay13}. The Ca/He and C/He abundances, calculated assuming $\log g=8.0$, are indicated by -log(Ca/He) or - log(C/He).}\\
21 & T\_err  (3D)& $T_\mathrm{eff}$ uncertainty \\
22 & log\_g  (3D)& $\log{g}$ \\
23 & log\_gerr  (3D)& $\log{g}$ uncertainty\\
24 & Mass & calculated mass ($M_\odot$) \\
25 & Mass\_err & mass uncertainty ($M_\odot$)\\
\end{tabular}
\end{table*}
\section{Conclusions and Discussion}

We have identified $9\,089$ new white dwarf and subdwarf stars in the DR 10 of the SDSS,
and estimated the mean masses for DAs and DBs, as well as the calcium contamination in DZs and carbon in DQs.
We were able to extend our identifications down to $T_\mathrm{eff}=5\,000$~K, although certainly not
complete, as we relied also on proper motion measurements, which are incomplete below $g\simeq 21$.
The substantial increase in the number of spectroscopically confirmed white dwarfs is important because it
allows the discovery of the rare objects, like the massive white dwarfs and oxygen spectrum over helium dominated atmosphere ones.
The improvements in the signal to noise and spectral coverage of BOSS spectra versus the SDSS spectrograph
also allowed for improvements in the stellar parameters, though systematic effects mainly due to flux calibration
must be explored.

\section*{Acknowledgments}
S.O. Kepler, 
I. Pelisoli, 
G. Ourique,
A. D. Romero,
and 
J.E.S. Costa
are sup\-por\-ted by CNPq and FAPERGS-Pronex-Brazil.
DK received support from programme Science without Borders, MCIT/MEC-Brazil.
BK is supported by the MICINN grant AYA08-1839/ESP, by the ESF EUROCORES
Program EuroGENESIS (MICINN grant EUI2009-04170), by the 2009SGR315 of the
Generalitat de Catalunya and EU-FEDER funds.
ARM acknowledges financial support from the Postdoctoral Science Foundation of China (grants 2013M530470 and 2014T70010) 
and from the Research Fund for International Young Scientists by the National Natural Science Foundation of China (grant 11350110496).

Funding for SDSS-III has been provided by the Alfred P. Sloan Foundation, the Participating Institutions, 
the National Science Foundation, and the U.S. Department of Energy Office of Science. The SDSS-III web site is http://www.sdss3.org/.

SDSS-III is managed by the Astrophysical Research Consortium for the Participating Institutions of the SDSS-III Collaboration including the University of Arizona, the Brazilian Participation Group, Brookhaven National Laboratory, Carnegie Mellon University, University of Florida, the French Participation Group, the German Participation Group, Harvard University, the Instituto de Astrofisica de Canarias, the Michigan State/Notre Dame/JINA Participation Group, Johns Hopkins University, Lawrence Berkeley National Laboratory, Max Planck Institute for Astrophysics, Max Planck Institute for Extraterrestrial Physics, New Mexico State University, New York University, Ohio State University, Pennsylvania State University, University of Portsmouth, Princeton University, the Spanish Participation Group, University of Tokyo, University of Utah, Vanderbilt University, University of Virginia, University of Washington, and Yale University.

\begin{table*}
\tiny
\noindent
\begin{minipage}{\linewidth}
\label{dados}
\caption{New White Dwarf Stars. Notes: 
P-M-F are the Plate-Modified Julian Date-Fiber number that designates an SDSS spectrum.
A {\bf:} designates an
uncertain classification. 
The columns are fully explained in Table~\ref{tb:columns}.
When $\sigma(\log g)=0.000$, we have assumed $\log g=8.0$, not fitted the surface gravity.
{\bf The full table is available on http://astro.if.ufrgs.br/keplerDR10.html}.}
\tiny
\begin{verbatim}
#SDSS J             Plate-MJD-Fiber S/N  u    su    g    sg     r    sr    i     si    z     si    ppm   Teff  sTeff logg  slogg Type T(3D) sT  logg   slogg Mass  sMass
#                                       (mag) (mag) (mag) (mag) (mag) (mag) (mag) (mag) (mag) (mag) 0.01"  (K)   (K)  (cgs) (cgs)       (K)  (K) (cgs)  (cgs) (Msun) (Msun)
#                                                                                                   /yr
000111.66+000342.55 4216-55477-0816 019 19.21 00.03 19.26 00.03 19.33 00.02 19.39 00.03 19.55 00.06 00.00 11323 00124 8.000 0.000 DB
000116.49+000204.45 4216-55477-0200 026 18.87 00.02 18.78 00.01 18.90 00.01 19.04 00.01 19.23 00.05 05.46 11107 00090 8.000 0.000 DBA
000216.03+120309.36 5649-55912-0467 002 22.88 00.43 22.02 00.08 21.97 00.11 22.20 00.19 22.38 00.61 00.00 10464 00433 8.584 0.212 DA  10422 00433 8.310 0.210 0.787 0.130
000243.16+073856.25 4535-55860-0842 045 17.67 00.01 17.52 00.01 17.86 00.01 18.11 00.01 18.38 00.02 01.41 20153 00080 8.178 0.001 DA  20153 00080 8.180 0.000 0.716 0.000
000247.21+101144.16 4534-55863-0584 013 20.84 00.07 19.71 00.02 19.00 00.02 18.88 00.02 18.76 00.04 00.00 06302 00006 5.430 0.240 DAZ:
000257.88+114719.07 5649-55912-0400 001 23.05 00.53 22.21 00.09 21.94 00.12 22.32 00.22 22.57 00.65 00.00 12628 01528 9.380 0.400 DA  12862 01528 9.190 0.400 1.237 0.136
000307.89+111732.42 5649-55912-0384 002 22.71 00.37 21.95 00.08 21.79 00.11 21.93 00.16 21.04 00.27 00.00 10235 00353 8.377 0.352 DA  10190 00353 8.100 0.350 0.657 0.190
000310.38+071801.11 4535-55860-0159 030 18.11 00.01 18.28 00.01 18.68 00.01 18.96 00.01 19.24 00.06 01.89 30455 00130 8.116 0.003 DA  30455 00130 8.120 0.000 0.705 0.000
000316.36+133829.22 5649-55912-0584 001 23.03 00.98 22.20 00.07 22.21 00.13 22.62 00.30 21.96 00.52 00.00 11045 00636 7.932 0.429 DA  11136 00636 7.740 0.430 0.490 0.180
000321.60-015310.86 4365-55539-0502 015 19.24 00.03 19.19 00.01 19.30 00.01 19.48 00.02 19.71 00.07 05.00 06039 01000 7.984 0.431 DA-He 06050 01000 7.990 0.430 0.584 0.225
\end{verbatim}
\end{minipage}
\end{table*}
\label{lastpage}
\end{document}